\documentstyle[11pt,epsfig]{article}

\title{\bf Correlations between multiplicities and average
transverse momentum in the percolating colour strings approach}

\author{M.A.Braun$^{a,b}$, R.S.Kolevatov$^a$,
C.Pajares$^b$ and  V.V.Vechernin$^a$\\
$^a$)Dep. High-Enegy Physics
S.Petersburg University, 198504 S.Petersburg, Russia\\
$^b$)Dep. Particle Physics, University of Santiago de Compostela,\\
15704 Santiago de Compostela, Spain}
\date{}
\def\beq{\begin{equation}}
\def\eeq{\end{equation}}

\begin{document}
\maketitle
\medskip
\vspace{1 cm}

\begin{abstract}
Long range correlations multiplicity-multiplicity, $p_T^2$-multiplicity
and $p^2 - p^2_T$ are studied in the percolating colour string picture
under different assumptions of the dynamics of string interaction.
It is found that the strength of these correlations is rather insensitive
to these
assumptions nor to the geometry of formed fused string clusters.
Both multiplicity-multiplicity and $p_T^2$-multiplicity correlations
are found to scale and depend only on the string density.
$p_T^2$-multiplicity correlations, which are absent in the independent
string picture, are found to be of the order of 10\% for central
heavy ion collisions and can
serve as a clear signature of string fusion.
In contrast
$p^2_T - p^2_T$ correlations turned out to be inversely proportional to
the number of strings and so very small for relaistic  collisions.
\end{abstract}
\section{Introduction}
The standard description of particle production in soft high-energy strong
interactions  is in terms of colour strings stretched between the
participant hadrons, whose decay generates the observed particle
spectrum. The number of strings grows with energy and atomic number of
participants and reaches thousands in heavy ion collision at RHIC and
LHC energies. With their density high enough, one expects that the
strings begin to overlap and interact. Some time ago a scenario of this
interaction was
introduced  based on the percolation phase transition
and existence of colour strings of high colour [1]. An immediate consequence
of the percolationg string scenario is damping of the multiplicity and
rising of the average transverse momentum as compared to the independent
string picture. These consequences are in agreement with the first data
obtained at RHIC [2]. It is to be stresed that within the percolating
string scenario one may imagine different versions of
of the string interaction dynamics. Overlapping strings may
not interact at all. Then the consequence of overlapping will be just
formation of different spots in the interaction area with higher
colour content corresponding to the sum of colours of the
overalpping strings. These spots will act as independent particle emitters
("overlaps, or "ministrings"). The number of ministrings will generally
be much greater
than that of initial strings. So in this scenario percolation and fusion
of strings actually leads to a proliferation of particle emitters.
An opposite scenario with a lot of interaction of overlapping strings
assumes that their colour becomes homogeneously distributed over the
formed cluster, the latter as a whole becoming an independent particle
emitter. In this scenario the number of particle emitters is evidently
smaller than the number of initial strings and may  become unity
if all strings form a single cluster. It is remarkable that these two
scenarios actually lead to practically the same predictions for such
global quantities as the multiplicity and average transverse momentum.
So to distinguish between them one has to study  more detailed
information about particle spectra. Immediate candidates are the
long range correlations between multiplicities or multiplicities and average
transverse momentum. In this note we study these observables in both
discussed scenarios of string interaction. Our results show that
unfortunately these long range correlations
are not very sensitive to the choice of string dynamics either.
However there is one advantage of studying the long-range correlations
involving the average transverse momenta: these correlations are
absent in the independent string model. So there presence is a
clear signature of string fusion and appearance of spots with higher
colour density than on the average. As we shall see, the correlations
between the  transverse momentum and multiplicity are of especial
value, since they depend only on the string density and not on
the total number of strings. Long range correlations between transverse
momenta, on the contrary, are inversely proportional to the total
number of strings and so are practically absent in central heavy ion
collisions. They can only be expected to be visible in highly
peripheral collisions or collisions involving light nuclei.

Some comments are to be made to explain the application of the
colour string approach to correlations. First we shall study
long-range correlations, which correspond to choosing the
correlating observables from different rapidity windows separated
by a reasonably large rapidity interval. This is to exclude short
range correlations which are always present and appear both at
the string decay stage (internal correlations
of string particle production) and final hadronization stage
(resonance decays). Second, in the study of average transverse momentum
we assume the chosen rapidity windows to exclude the fragmentation
regions, where the correlations between the average transverse momentum
and multiplicities may largely be a result of summing the transverse
momenta of the parent partons from the colliding hadrons. In fact the
latter explanantion is the standard one for the observed correlations [3].
However it does not apply to the central region, in which particle
production becomes insensitive to the parent parton momenta.  So hopefully
the correlations in the central region result only due to
string interactions.  Finally we have to remember that the
unperturbative colour string picture in the present version is oriented only
to soft particle spectra. So our region of average transverse momenta has to
be limited by values of the order of 1$\div$ 2 GeV/c.

\section{General formalism}

In this subsection we shall present our basic probabilistic expressions
for the correlations valid for any scenario of string interaction.

In a given event colour strings occupy certain regions in the transverse
interaction plane and may overlap forming clusters. In any scenario
this generates independent emitters, either overlaps or clusters as a whole.
We enumerate these emitters by subindex $\alpha=1,2,3,...M$,
where $M$ is the total number of emitters.

An event consists of the emission of $m_{\alpha}$ particles
from emitter $\alpha$. The total number of emitted particles is
\beq
m_e=\sum_{\alpha}m_{\alpha}
\eeq
The $p_T$ distribution of particles emitted from emitter $\alpha$ will
be given by a certain function $w_\alpha(p)$. Note that
it corresponds to the average
\beq
p_{\alpha}^2=\int d^2p p^2w_\alpha(p)
\eeq
whose concrete value depends of the scenario for the interaction.
The effective $p_T$ distribution in a given event is then
\beq
w_e(p)=\frac{1}{m_e}\sum_{\alpha}m_{\alpha}w_\alpha(p)
\eeq
and the average $p_T^2$ is
\beq
p_e^2=\frac{1}{m_e}\sum_{\alpha}m_{\alpha}p_\alpha^2
\eeq

Now we pass to taking averages over many events. This averaging
may be divided in two steps. First we average over various
events occurring with the same string geometry,
that is, with the fixed overlap and cluster structure ("configuration").
Then we have to average over all possible configurations.

To average over events at a fixed configuration we have to know
the probability $\rho_{\alpha}(m)$ for a given emitter to produce
$m$ particles. This probabilty has to lead to the average number of particles
\beq
\bar{m}_{\alpha}=\sum_m m\rho_{\alpha}(m)
\eeq
which again is to be taken according to the chosen scenario of string
interaction.
We get averages in a given configuration:
\beq
\bar{m}_c=\sum_{\{m_{\alpha}\}}\prod_{\alpha}\rho_{\alpha}(m_{\alpha})m_e=
\sum_{\alpha}\bar{m}_{\alpha}
\eeq
\beq
\bar{p^2}_c=\sum_{\{m_{\alpha}\}}\prod_{\alpha}\rho_{\alpha}(m_{\alpha})
\frac{1}{m_e}
\sum_{\alpha}m_{\alpha}p^2_{\alpha}
\eeq
and also the probabilities in a given configuration to emit $m$ particles
\beq
\rho_c(m)=\sum_{\{m_{\alpha}\}}\prod_{\alpha}\rho_{\alpha}(m_{\alpha})
\delta_{m,\sum_{\alpha}m_{\alpha}}
\eeq
and a particle with the transverse momentum $p$
\beq
w_c(p)=\sum_{\{m_{\alpha}\}}\prod_{\alpha}\rho_{\alpha}(m_{\alpha})
\frac{\sum_{\alpha}m_{\alpha}w_{\alpha}(p)}{\sum_{\alpha}m_{\alpha}}
\eeq

To study correlations we have to know double distributions.
We assume that emissions in the forward  and backward
rapidity windows are independent. As a result
the double distribution in the numbers of particles $m_F$ and $m_B$
emitted in the forward and backward rapidity windows in a given
configuration factorizes
\beq
\rho_c(m_F,m_B)=\rho^F_c(m_F)\rho^B_c(m_B)
\eeq
where $\rho^{F,B}_c(m)$ are given by (8) with emitter probabilities
$\rho^{F,B}_{\alpha}(m)$ to produce particles in the forward (F) or
backward (B) rapidity windows.
Similarly the double distribution in the number of particvles $m_F$ in the
forward rapidity window and the particle transverse momentum $p_B$ in the
backward rapidity window in a given configuration is a product
\beq
\rho_c(m_F, p_B)=\rho_c^F(m_F)w^B_c(p_B)
\eeq
Here $w^B(p)$ is given by (9) with $\rho_{\alpha}\to \rho^B_{\alpha}$.
Finally the double distribution in momenta $p_F$ and $p_B$ of particles emitted
in the forward and backward rapidity windows  factorizes   as
\beq
w_c(p_F, p_B)=w_c^F(p_F)w^B_c(p_B)
\eeq

To average over different configurations we have to sum these expressions
over all configurations with the weight $P_c$, which is the probability
of a given configuration. We denote this operation as $<\ >$.
In particular we find various conditional probabilities of interest.
The conditional probability to find $m_B$ particles in the backward rapidity
window provided one sees $m_F$ particles in the forward rapidity window is
\beq
<\rho(m_B)>_{m_F}=\frac{\langle \rho_c(m_F,m_B)\rangle}
{\langle \rho_c^F(m_F)\rangle}
\eeq
The conditional probability to find a particle with transverse momentum
$p_B$ in the backward rapidity window provided one sees
$m_F$ particles in the forward rapidity window is
\beq
<w(p_B)>_{m_F}=\frac{\langle \rho_c(m_F,p_B)\rangle}
{\langle \rho_c^F(m_F)\rangle}
\eeq
Finally the conditional probability to find a particle with momentum
$p_B$ in the backward rapidity window provided one sees a particle with
momentum $p_F$ in the forward rapidity window is
\beq
<w(p_B>_{p_F}=\frac{\langle w_c(p_F,p_B)\rangle}
{\langle w_c^F(p_F)\rangle}
\eeq

Taking the averages with these probabilities we find our basic
formulas for correlations.
The average multiplicity in the backward window at a given multiplicity
in the forward window is
\beq
<m_B>_{m_F}=\frac{\langle (\bar{m_B})_c\rho_c^F(m_F)\rangle}
{\langle \rho_c^F(m_F)\rangle}
\eeq
where $(\bar{m}_B)_c$ and $\rho^F(m_F)$ are given by (6) and (8) for the two
rapidity windows.
The average transverse momentum squared in some rapidity window at a
given multiplicity in a different rapidity window is
\beq
<{p_B}^2>_{m_F}=\frac{\langle (\bar{{p_B}^2})_c\rho_c^F(m_F)\rangle}
{\langle \rho_c^F(m_F)\rangle}
\eeq
where $(\bar{{p_B}^2})_c$ is given by (7) with $\rho_{\alpha}\to
\rho^B_{\alpha}$.
Finally the average transverse momentum squared in some rapidity window at a
given momentum in a different rapidity window is
\beq
<{p_B}^2>_{p_F}=\frac{\langle (\bar{{p_B}^2})_cw^F_c(p_F)\rangle}
{\langle w^F_c(p_F)\rangle}
\eeq
where $w^F_c$ is given by (9) with $\rho_{\alpha}\to \rho^F_{\alpha}$
These formulas serve as a starting point for our study.

Expressions for the correlations can be substantially simplified if the
emission probabilities $\rho_{\alpha}(m)$ have the Poisson form
\beq
\rho_{\alpha}(m)=P_{\bar{m}_{\alpha}}(m)=
e^{-\bar{m}_{\alpha}}\frac{\bar{m}_{\alpha}^m}{m!}
\eeq
Then taking average in a given configuration can be done
analytically.

In particular we find with (19) (see Appendix 1.)
\beq
\rho_c(m)=P_{\bar{m}_c}(m)=e^{-\bar{m}_c}\frac{\bar{m}_c^m}{m!}
\eeq
\beq
\bar{p^2}_c=\frac{1}{\bar{m}_c}\sum_{\alpha}\bar{m}_{\alpha}p^2_{\alpha}
\eeq
and
\beq
w_c(p)=\frac{1}{\bar{m}_c}\sum_{\alpha}\bar{m}_{\alpha}w_{\alpha}(p)
\eeq
For particles emitted in the forward (backward) rapidity windows
one has to substitute in these formulas $\bar{m}\to
\bar{m}_{F,B}$

With this simplification, calculation of
correllations reduces to taking the averages only over configurations.

\section{Multiplicity correlations}
Forward-backward (FB) correlations between multiplicities has long
been studied both experimentally [4] and theoretically [5--7]. On the
theoretical level they demonstrate that the number of elementary
emitters fluctuates. On this ground they have served as a confirmation
of the colour string picture, showing that the number of strings
fluctuates, with its average  growing with energy. For the
FB multiplicity correlations  to exist, the interaction between strings
is irrelevant: the correlations are fully present when strings are
independent.
Moreover, as conjectured in [8], interaction and fusion of strings lead
to damping of these correlations, since the effective number of
strings diminishes. As we shall see this conjecture is confirmed by
calculations using Eq. (16) in any of the two scenarios for the string
interaction.

\subsection{Ministrings as emitters}
As mentioned in the Introduction the two scenarios differ in the
intensity of the interaction between the overlapping strings.
The ministring scenario assumes that this interaction is in fact
absent: the overlap regions retain their form and serve as
independent emitters of particle with characteristics determined
by the total colour accumulated in the overlap. In this scenario
different emitters can be labeled by two numbers $\alpha=\{ni\}$ where
$i=1,2,...$ enumerates overlaps of $n$ strings. The basic quantity which
characterizes the ministring scenario is the average multiplicity of
overlap $ \{ni\}$ [4,5]:
\beq
\bar{m}_{ni}=\mu_0\sqrt{n}\frac{S_{ni}}{\sigma_0}
\eeq
Here $S_{ni}$ and $\sigma_0$ are transverse areas of the overlap and
initial string respectively. Using (25) we obtain the average
multiplicity in a given configuration as
\beq
\bar{m}_c=\mu_0\sum_{n}\sqrt{n}\frac{S_n}{\sigma_0}
\eeq
where $S_n=\sum_iS_{ni}$ is the total area  in which $n$ strings overlap
and $\mu_0$ is the average multiplicity for the initial string.
The average multiplicities in the forward and backward rapidity
windows are given by the same formula with $\mu_0\to\mu_0^{F,B}$,
the forward and backward multiplicities for the initial string:
\beq
\bar{m}^{F,B}_c=\mu^{F,B}_0\sum_{n}\sqrt{n}\frac{S_n}{\sigma_0}
\eeq
Putting (25) into (20) and using (16) we reduce calculation of
the FB multiplicity correlations to averaging over the overlap
geometries of the known expression depending on $S_n$. This averaging
involves, first, different ways in which string overlap with their total
$N$ number fixed and, second, different values of $N$ distributed around
a certain average value $<N>$ determined by the energy and atomic number
of participants. At large $<N>$ and small $\sigma_0$ as compared to the
total interaction area $S$ one expects a scaling behaviour:
appropriately chosen characteristics of the correlations
depend only on the dimensionless percolation parameter
\beq
\eta=\frac{<N>\sigma_0}{S}
\eeq
This property allows to study the correlations by Monte-carlo simulations
with comparatively low values of $<N>$. Values of $\eta$ and $<N>$ for S-S
and Pb-Pb
central collisions at different energies obtained by Monte Carlo
simulations based on the fusing colour string model  [1] are
shown in Table 1.

In our numerical caluclations we studied the quantity
\beq
F_{\mu-\mu}=\frac{\langle m_B\rangle_{m^F}}{\langle m_B\rangle}-1
\eeq
as a function of
\beq
x=\frac{m_F}{\langle{m_F}\rangle}\eeq
In our simulations we used both the homogeneous distribution of strings
in the transverse interaction area and the inhomogeneous one, which
follows the nuclear profile function $T(b)$. No significant qualitative
difference was found between these two choices. For simplicity we
present here our results for the homogeneous distribution.
We chose $\mu_0^{F,B}=1$.
We used the Poisson distribution in the
number of strings $N$ with $<N>=25$ and 50. We have checked that the results
are practically independent of $<N>$ in the
region $0.3<x<3$, so that $F_{\mu-\mu}(x)$ exhibits the
expected scaling behaviour for physically relevant values of $m_F$.

Our results for values of $\eta=0.5\div 3$ show that in the whole
interval $0<x<3$
$F_{\mu-\mu}(x)$ is  a
monotonously rising function of $x$, rather close to linear and crossing
zero at $x=1$.  Its slope at $x=1$ (and $<N>=50$)
(the "correlation coefficient")
\[
b_{\mu-\mu}=\left(\frac{dF_{\mu-\mu}(x)}{dx}\right)_{x=1}
\]
is shown in Fig. 1 by a solid line.
As we observe it steadily falls with $\eta$, in full accordnace with the
idea that the
correlations diminish as a consequence of string fusion. At very high
$\eta$ the slope seems to flatten against the value around 0.1.
At $\eta\to 0$
the slope tends to the value 0.5, which corresponds to the independent string
model with the assumed value of $\mu_0$ (see Appendix 1).
Note that
$\eta\sim 1.12$ corresponds to the critical value for the percolation
phase
transition. However our curves do not show any peculiarity at such values of
$\eta$.

In Fig. 2 we compare $b_{\mu -\mu}$ for the homogeneous distribution of
strings and the inhomogeneous one which follows the nuclear ptofile
function $T(b)\propto \sqrt{R_A^2-b^2}$ (corresponding to a constant
nuclear density). As mentioned, the difference is of no practical
importance.
\subsection{Clusters as emitters}
An alternative scenario with a strong interaction between overlapping strings
assumes that the colour becomes homogeneously distributed over the whole
cluster formed by the overlapping strings. The emitters are now clusters
themselves  and index $\alpha$ now labels just different clusters.
The average multiplicity of a cluster  is chosen to adjust to the
colour distribution [9]:
\beq
\bar{m}_{\alpha}=\mu_0\sqrt{\frac{n_{\alpha}S_{\alpha}}{\sigma_0}}
\eeq
where $n_{\alpha}$ and $S_{\alpha}$ are the number of strings and area
of cluster $\alpha$. For the forward and backward rapidity window
(29) transforms into
\beq
\bar{m}^{F,B}_{\alpha}=\mu^{F,B}_0\
\sqrt{\frac{n_{\alpha}S_{\alpha}}{\sigma_0}}
\eeq
Eqs. (29) and (30) express the physical contents of the cluster scenario for
multiplicity correlations and their difference from Eqs. (24) and (25)
shows the effect of the strong interaction between overlapping strings.
The remaining procedure to calculate the correlations does not change
as compared to the ministring case.

We calculated the same function $F_{\mu-\mu}(x)$ (Eq. (29))
with (32) by Monte Carlo
simulations. Comparison of our results for $<N>=25$ and 50 again
showed a satisfactory scaling in $<N>$ for $x=0.3\div 3$ and a weak
dependence on the way the strings are distributed in the interaction area.
In spite of a very different dynamics, we have found no appreciable
difference with the former case ov overlaps as emitters.
The slopes $b_{\mu-\mu}$ of $F_{\mu-\mu}(x)$ at $x=1$ are shown in
Fig. 1 by a
long-dashed line (again for $<N>=50$). They are systematically higher
than for overlaps but the difference does not exceed 10\%, which is
of no practical importance, taking into account simplifications
involved at the basis  of the model.

\subsection{Fixed number of emitters}
Technically taking the average over all geometrical distributions
of strings in the transverse area is a formidable task, which can be
realistically achieved by Monte Carlo simulations for a reasonable time
only for a limited number of strings, substantially smaller than this number
in heavy-ion collisions at RHIC and especially LHC energies. On the
other hand, it is well-known that different ways of averaging often
lead to the same or practically the same averages, since the physical
picture of fluctuating variables is often basically the same.
This motivates searching for a simplified picture of string fusion,
which, on the one hand, is not very different from the discussed above
by its physical implications and, on the other hand, avoids
geometrical averaging.   In this subsection we propose  such a
picture.

Physically the basis of  string fusion model  consists of appearance
of various emitters homogeneously distributed in the transverse plane
with the number of overlapping strings varying for 1 to $N$ with
a certain probability. One can model this physical situation
by assuming a fixed number $M$ of emitters {"cells"), each one of them
(emitter $\alpha$)
equivalent to a certain number $n_{\alpha}$
of completely overlapped strings
including $n_{\alpha}=0$, in which case there is no emission at all.
So effectively one also has a varying number of emitters with different
colours. Averaging over configurations is achieved by distributing
each $n_{\alpha}$ around some average value $\bar{n}_{\alpha}$ with
a certain probability $P_{\alpha}(n_\alpha)$. The homogeneous
distribution of strings in the transverse plane corresponds to equal
averages: $\bar{n}_{\alpha}=\bar{n}$ . One can model an
inhomogeneous distribution of strings by distributing the $M$ emitters
in the transverse plane and choosing $\bar{n}_{\alpha}$ in accordance with
the nuclear profile density.

In this model ("cell model" [10]) we have for the average number of particles
from cell $\alpha$
\beq
\bar{m}_{\alpha}=\mu_0\sqrt{n_{\alpha}},\ \
\bar{m}^{F,B}_{\alpha}=\mu^{F,B}_0\sqrt{n_{\alpha}}
\eeq
If we assume that the cell probabilities $\rho_{\alpha}(m)$ are Poissonian
then correlations can be calculated using our formulas of Section 2 in
exactly the same way as before. For the final averaging over configurations
(that is over all $n_{\alpha}$, $\alpha=1,...M$) we have taken the
probabilities $P_{\alpha}(n_\alpha)$  to be also Poissonian with
a common average value $\bar{n}=\eta$ to correspond to the previous
geometrical picture. Note that in this model the average value of strings is
$<N>=M\eta$.

The results of calculations of $F_{\mu-\mu}(x)$ in this model for
$<N>=50$ and various $\eta$ show that they are very close to both
previous models, especially at relatively large $\eta$ when they are
nearly identical to those in the ministring picture.
The slopes of $F_{\mu-\mu}(x)$ at $x=1$ for the cell model are shown in
Fig. 1 with a short-dashed line.
So as expected, this simplified model imitates the geometrical model
of ministrings almost ideally. With that, it is much simpler: its
realization by Monte Carlo simulations requires
computer time more than an order of magnitude smaller than for ministrings
with equal $<N>$. Apart from this, the model admits some analytic estimates
for small or large values of $\eta$, which reveal basic properties
of the correlaltions in these asymptotical regions (see Appendix 2.).

\section{Correlations between average transverse momentum and
multiplicities}
According to Eq. (17) to calculate the correlations $<p^2_T>-\mu$
we have to know the averages $\bar{p^2}_c$ in the backward window
in a given configuration. For the Poissonian emitters they can be calculated
via Eq. (21) if one knows these averages for a given emitter $\alpha$
(Eq. (2)). The concrete values of $p^2_{\alpha}$ depend on the
scenario of string interaction. They were analyzed in [8,9] wherefrom we
borrow their expressions presented below.

Both for the ministring and cell scenarios the value of $p^2_{\alpha}$
for the individual emitter are determined only by the the number of
strings in the overlap or cell. For ministrings
\beq
p^2_{ni}=p_0^2\sqrt{n}
\eeq
and for the cell model
\beq
p^2_{\alpha}=p_0^2\sqrt{n_\alpha}
\eeq
For the cluster scenario $p^2_\alpha$ depends on the color density inside
the cluster and thus also depends on its area $S_{\alpha}$:
\beq
p^2_{\alpha}=p_0^2\sqrt{\frac{n_{\alpha}\sigma_0}{S_{\alpha}}}
\eeq
In these formulas $p_0^2$ is the average transverse momentum squared for the
initial string.

With these expressions for $p_{\alpha}^2$ we calculated
$\langle p_B^2\rangle_{m_F}$ according to Eq. (17) by Monte Carlo
simulations for the three scenarios considered: ministrings, clusters and
cells.  We studied
\beq
F_{p^2-\mu}=\frac{\langle p^2_B\rangle_{m^F}}{\langle p^2_B\rangle}-1
\eeq considered as a function of the same variable $x$, Eq.(28).
As before we took $<N>=25$ and 50  and  $\eta$ in the region $0.5\div 3$.
We found that  $F_{p^2-\mu}(x)$
again exhibits a satisfactory scaling property: it results practically
independent of $<N>$.
As for $\mu-\mu$ correlations $F_{p^2-\mu}(x)$  is found to be a
nearly linearly rising function of $x$, crossing zero at $x=1$.
However, contrary to $\mu-\mu$ correlations, the slopes
for $p^2-\mu$ correlations rise with $\eta$. Slopes of
$F_{p^2-\mu}(x)$ at $x=1$
\[b_{p^2-\mu}=\left(\frac{dF_{p^2-\mu}(x)}{dx}\right)_{x=1}\]
are shown in Fig. 3 for the three studied
scenarios with the homogeneous distribution of strings. One observes that
the overlap and cluster scenarios
again lead to practically identically results. The cell model imitates these
two physical scenarios quite satisfactory at large $\eta>2\div 2.5$.
At smaller $\eta$ its slopes lie 30$\div$40\% below. At large $\eta$
slopes in all three models seem to converge around 0.1. It is remarkable
that this value coinsides with the corresponding limit for the
slopes in $\mu-\mu$ correlatioms (cf. Fig. 1).  At $\eta\to 0$ all
slopes vanish, which corresponds to absence of $p^2-\mu$ correlations
for independent strings.

In Fig. 4 we compare the slopes in the overlap scenario for the
homogeneous distribution of strings and the same realsitic one,
as in Fig. 2. Again the difference results of little importance.

It has to be stressed that absolute values of the slopes are not too small,
of the order of 0.1 in a wide range of $\eta$. Such slopes can hopefully
be measured experimentally. There existence would be a clear signature
of string fusion.

\section{$p_B-p_F$ correlations}
The last type of correlations we shall study are those between transverse
momenta of particles emitted in the forward and backward  rapidity windows.
They can be measured by the average transverse momenta squared of particles
in the backward rapidity window at an observed transverse momentum
in the forward rapidity window, Eq. (18). To calculate this in our
colour string model we have to know the particle spectrum $w_{\alpha}(p)$
for an individual emitter. This spectrum has to be chosen to
lead to the averages (2), which have been discussed for each of
the models for string interaction in the previous section.
We choose the distribution $w_{\alpha}(p)$ in the form prompted by
the standard fit to the experimental spectrum for the proton target [11]
\beq
w_{\alpha}(p)=\frac{(\kappa-1)(\kappa-2)}{2\pi p_{\alpha}^2}
\left(\frac{p_{\alpha}}{p+p_{\alpha}}\right)^{\kappa}
\eeq
Here $p^2_{\alpha}$ is given by Eqs. (32)-(34) for the three models
of string interaction considered, $p_{\alpha}=\sqrt{p^2_{\alpha}}$,
\beq
 \kappa=19.7-0.86\ln E,\ \ p_0^2=\frac{24}{(\kappa-3)(\kappa-4)}\ \
 {\rm (GeV/c)}^2
\eeq
and $E$ is the c.m. energy in GeV.

With this distribution we calculated (18) for the three considered scenarios
averaging over configurations by Monte Carlo simulations with different
average number of strings $<N>=25$ and 50. Calculations show that for
these correlations the scaling function independent of $<N>$ turns
out to be
\beq F_{p-p}=<N>\frac{<p_B^2>_{p_F}}{<p_B^2>}-1
\eeq
As argument it is convenient to choose dimensionless
\beq
y=\frac{p_F}{\sqrt{<p_F^2>}}
\eeq
We have calculated $F_{p-p}(y)$ for the
for different $\eta$.
In all cases $F_{p-p}(y)$ is found to be a linearly rising function of $y$
crossing zero at $y=1$. The slopes of $F_{p-p}(y)$ at $y=1$
\[ b_{p-p}=\left(\frac{dF_{p-p}(y)}{dy}\right)_{y=1}\]
are
shown in Fig. 5 as  functions of $\eta$  for the three studied models.
They rise with $\eta$ similarly to the slopes in $p^2-\mu$
correlations. Again overlaps and clusters lead to practically
identically results. Their imitation by cells gives rather satisfactory
results, especially at $\eta>2$ where the cell slopes lie higher
than those from the other two scenarios by some 10\%. At lower $\eta$ this
difference rises to 30$\div$40 \%.

Note that the presence of the factor $<N>$ in
(38) together with the calculated values of the scaling function
show that for
realistic collisions with high enough values of $\eta$ the magnitude
of the correlations is extremely small, which excludes their
experimental observation. As follows from our results the correlations
can be observed in practice up to $<N>=10\div 15$, which is
unrealistic for  $\eta>1$.

\section{Conclusions}
This work was primarily motivated by search of appropriate observables
which could shed light on the details of dynamics of colour string
fusion conjectured to occur in heavy-ion collisions. The results of
our study show that pair correlations are unfortunalely  rather
insensitive to these details. Moreover they demonstrate that even the
geometric picture of string fusion and percolation is not very
essential for the found correlations. A very simplified picture,
which only takes into account existence of many emitters with different
colours, leads to results not very different from the realistic
geometrical overlapping of colour strings. So the correlations
basically reflect only the possibility of spots with higher colour field
in the colliding nuclei and strongly depend only on their spatial density
characterized by the parameter $\eta$.

From a certain point of view this is a positive aspect, since predictions
for  the correlations do not vary  with specific assumptions made about
the string fusion dynamics. The  behaviour of the
correlations with $\eta$ emerges as a clear signature for the
basic new phenomenon characteristic for string fusion:
formation of emitters with higher colour density. Both $\mu-\mu$ and
$p^2 -\mu$ correlations look quite promising from this point of view.
In particular  $p^2-\mu$ correlations arise totally due to string
fusion in our picture; they are absent with independent strings.
However, as mentioned in the Introduction, this is only true for
particles produced strictly in the central region, since in the
fragmentation regions obvious $p^2-\mu$ correlations are caused by
the accumulation of the string ends momenta. So to see the effect of
string fusion one should choose both rapidity windows as far as
possible from the rapidity limits. As to $p-p$ correlations, our
calculations have shown that they diminish with the number of strings
and appear to be negligible for realistic heavy-ion collisions
except at the highly peripheral collisions.

Our calculations have been made for a homogeneous distribution of
strings in the transverse interaction plane. Calculations for
a realistic distribution, taking into account the varying string density
as a function of impact parameter, are now in progress. However
the found insensitiveness of our results to the details of geometry
allows to expect that the change introduced by a realitic geometry will
be not very essential.

\section{Acknowledgments}
The authors are deeply thankful to  Drs. G.A.Feofilov and E.G.Ferreiro
for a keen
interest in this work and helpful discussions.
This work was  supported by the RFFI (Russia) grant 01-02-17137
and contract FPA 2002- 01161 of CICYT (Spain)

\section{Appendix 1. Averages with the Poisson distribution}
We start with $\rho_c(m)$ determined by (8).
Presenting the Kronecker symbol as a contour integral around the origin we
find
\beq
\rho_c(m)=
\sum_{\{m_{\alpha}\}}\int \frac{dz}{2\pi iz}
z^{-m+\sum_{\alpha}m_{\alpha}}\prod_{\alpha}\rho_{\alpha}(m_{\alpha})
\eeq
We have
\beq
\sum_{m_{\alpha}}z^{m_{\alpha}}P(m_{\alpha})=e^{(z-1)\bar{m}_{\alpha}}
\eeq
so that
\beq
\rho_c(m)=\int\frac{dz}{2\pi z ^{m+1}}e^{(z-1)\bar{m}_c}=
e^{-\bar{m}_c}\frac{\bar{m}_c^m}{m!}
\eeq
So we find (20).

Now take average (7). We present it as
\beq
\bar{p^2}_c=\sum_m\frac{1}{m}
\sum_{\{m_{\alpha}\}}\prod_{\alpha}\rho_{\alpha}(m_{\alpha})
\delta_{m,\sum_{\alpha}m_{\alpha}}\sum_{\alpha}m_{\alpha}p^2_{\alpha}
\eeq
The internal sum over all $m_{\alpha}$ at their sum fixed can be
represented in the same way as (40):
\beq
X\equiv
\sum_{\{m_{\alpha}\}}\int \frac{dz}{2\pi iz}
z^{-m+\sum_{\alpha}m_{\alpha}}\prod_{\alpha}\rho_{\alpha}(m_{\alpha})
\sum_{\alpha}m_{\alpha}p^2_{\alpha}
\eeq
Take some particular term in the last sum, say $\alpha=1$
Then sums over all $m_{\alpha},\ \alpha=2,3,...M$ will give the
same factor (41). The sum over $m_{1}$ will however contain an extra
factor $m_1$:
\beq
\sum_{m_{1}}z^{m_{1}}m_1P(m_{1})=e^{(z-1)\bar{m}_{1}}=
z\bar{m}_1 e^{(z-1)\bar{m}_1}
\eeq
So we find
\beq
X=\sum_{\alpha}\bar{m}_{\alpha}p^2_{\alpha}e^{-\bar{m}_c}
\int \frac{dz}{z^{m_1}}e^{z\bar{m}_c}=
\frac{m}{\bar{m}_c}\rho_c(m)\sum_{\alpha}\bar{m}_{\alpha}p^2_{\alpha}
\eeq
where $\rho_c(m)$ is given by (20).
Putting this into (43) we find that the factor $m$ in (46) cancels
the denominator in (43). Using then
\[\sum_m\rho(m)=1\]
we find (21).

Derivation of (22) is done in exactly the same way.

\section{Appendix 2. Analytic estimates in the cell scenario}
As mentioned, the cell scenario admits explicit analytic estimates [10]
valid in the limit of  large or small values of $\eta$.

Estimates at large $\eta$ are based on the asymptotic equivalence
of the discrete Poisson distribution and continuous Gaussian
distribution
\beq
P_{\bar{n}}(n)\sim G_{\bar{n}}(n)=\frac{1}{\sqrt{2\pi\bar{n}}}
e^{-\frac{(n-\bar{n})^2}{2\bar{n}}}
\eeq
valid in the limit $\bar{n}>>1$.

Let us first study the simplest case of independent strings (without
fusion). Then Eqs. (31) and (33) are changed to
\beq
\bar{m}_{\alpha}=\mu_0 n_{\alpha},\ \ p^2_{\alpha}=p_0^2
\eeq
and similarly for averages in the two rapidity windows.
To simplify in this Appendix we shall take $\mu_0^F=\mu_0^B$ and denote
this common multiplicity by $\mu_0$. From the second of Eqs. (48) it follows
that there will be no $p^2-\mu$ correlations as expected.
The configuration averages $\bar{m}_c^{F,B}=\mu_0\sum_{\alpha}n_{\alpha}=
\mu_0N_c$ will depend only on the total number of strings $N_c$ in a
given configuration. Averaging over configuration will be reduced
to averaging over the number of strings in different configurations.
If the number of strings in each cell is distributed according
to the Poisson distribution with average $\eta$ then the total number
of strings will also be distributed according to the Poisson
distribition with the average $M\eta$ (see Appendix 1). So at fixed $\eta$
the overall average number of strings  is related to $M$ as
\beq
\langle N\rangle= M\eta
\eeq

Thus, substituting the two Poisson distributions by Gaussians we find
from (16)
\beq
\langle  m_B\rangle_{m_F}=\mu_0
\frac{\int_0^{\infty}NdN\,N^{-1/2}e^{-\phi(N,m_F)}}
{\int_0^{\infty}dN\,N^{-1/2}e^{-\phi(N,m_F)}}
\eeq
where
\beq
\phi(N,m_F)=\frac{(N-\langle N\rangle)^2}{2\langle N\rangle}
+\frac{(m_F-\mu_0N)^2}{2\mu_0N}
\eeq
We estimate the two integrals by the saddle point method.  Then
we find
\beq
\langle  m_B\rangle_{m_F}=\mu_0 N_0(m_F)
\eeq
where $N_0$ is the solution of the equation $d\phi/dN=0$.
In terms of scaling quantities
\beq
x=\frac{m_F}{\langle m_F\rangle},\ \  z=\frac{N}{\langle N\rangle}
\eeq
the equation for the saddle point takes the form
\beq
z^3-z^2=\frac{1}{2}\mu_0 (x^2- z^2)
\eeq
Solving this equation for $z_0(x)$ one can find function
$\langle  m_B\rangle_{m_F}(m_F)$ at all $m_F$.
The correlation coefficient $b_{\mu-\mu}$ can however be
determined explicitly from (54):
\beq
b_{\mu-\mu}=\left(\frac{dz}{dx}\right)_{x=1}=\frac{\mu_0}{\mu_0+1}
\eeq
Taking $\mu_0=1$ we have $b_{\mu=\mu}=0.5$ in full agreement
with the limit of curves shown in Fig. 1 at $\eta\to 0$, which
corresponds to independent strings.

Now we pass to a physically more interesting case when strings fuse,
which in the cell scenario corresponds to Eqs. (31) and (33).
Introducing a quantity  for a given configuration
\beq
r_c=\sum_{\alpha}\sqrt{n_{\alpha}}
\eeq
we find
\beq
\bar{m}_c=\mu_0 r,\ \ \bar{p^2}_c=p_0^2\frac{N_c}{r}
\eeq
so that
\beq
\langle m_B\rangle_{m_F}=\mu_0\frac{\langle rP_{\mu_0r_c}(m_f)\rangle}
{\langle P_{\mu_0r}(m_f)\rangle}
\eeq
and
\beq
\langle p^2_B\rangle_{m_F}=
p_0^2\frac{\langle (N_c/r_c)P_{\mu_0r}(m_f)\rangle}
{\langle P_{\mu_0r}(m_f)\rangle}
\eeq
The average is taken over the product of Poisson distributions in
$n_{\alpha}$ each with with the average $\eta$.

We consider $\eta>>1$ and substitute all
 Poisson distributions with Gaussian ones to get
\beq
\langle m_B\rangle_{m_F}=\mu_0
\frac{\int_0^{\infty}\prod_{\alpha}dn_{\alpha}
 r_cr_c^{-1/2}e^{-\phi(n_{\alpha},m_F)}}
{\int_0^{\infty}\prod_{\alpha}dn_{\alpha}
 r_c^{-1/2}e^{-\phi(n_{\alpha},m_F)}}
\eeq
and
\beq
\langle p^2_B\rangle_{m_F}=p_0^2\frac
{\int_0^{\infty}\prod_{\alpha}dn_{\alpha}
(N_c/r_c) r_c^{-1/2}e^{-\phi(n_{\alpha},m_F)}}
{\int_0^{\infty}\prod_{\alpha}dn_{\alpha}
 r_c^{-1/2}e^{-\phi(n_{\alpha},m_F)}}
\eeq
where
\beq
\phi(n_{\alpha},m_F)=\sum_{\alpha}\frac{(n_{\alpha}-\eta)^2}{2\eta}+
\frac{(m_F-\mu_0r_c)^2}{2\mu_0r_c}
\eeq

In the saddle point approximation we find
\beq
\langle m_B\rangle_{m_F}=\mu_0r_0,\ \ \
\langle p^2_B\rangle_{m_F}=p_0^2\frac{N_0}{r_0}
\eeq
where $N_0$ and $r_0$ are values of $N_c$ and $r_c$ at the saddle point
defined by the equations
\beq
\frac{\partial\phi(n_{\alpha}, m_F)}{\partial n_{\alpha}}=0,\ \ \alpha=1,...M
\eeq
Introducing scaled variables
\beq
x=\frac{m_F}{\langle m_F\rangle},\ \ z_{\alpha}=\sqrt{\frac{n_{\alpha}}
{\eta}},
\eeq
and a constant involving $\eta$
\beq
a=\frac{\mu_0}{4\sqrt{\eta}}
\eeq
we can write the saddle point equations in the form
\beq
z_{\alpha}^3-z_{\alpha}=a\left(x^2\frac{M^2\eta}{r_c^2}-1\right),
\ \ \alpha=1,...M
\eeq
In terms of $z_{\alpha}$ here $r_c=\sqrt{\eta}\sum_{\alpha}z_{\alpha}$.
Due to symmetry in $\alpha$ obviously for the solution
$z_{\alpha}=z$ where $z$ satisfies a single equation
\beq
z^3-z=a\left(\frac{x^2}{z^2}-1\right)
\eeq
This equation defines $z=z(x)$, in terms of which one finds
\beq
F_{\mu-\mu}(x)=F_{p^2-\mu}=z(x)-1
\eeq
So at large values of $\eta$  both $\mu-\mu$ and $p^2-\mu$ correlations
are described by the same scaling function of $x$.

At $x=1$ Eq. (68)  possesses an obvious trivial solution:
\beq
x=1,\ \ z_=1,\ \, r_0=M\sqrt{\eta},/ /  N_0=\langle N\rangle=\mu_0 M
\eeq
This is sufficient to calculate the (identical) correlation coeficients for
both $\mu-\mu$ and $p^2-\mu$ correlations:
\beq
b_{\mu-\mu}=b_{p^2-\mu}=\frac{\mu_0}{\mu_0+4\sqrt{\eta}}
\eeq
It slowly falls with $\eta$
Of course one should have in mind that thexe are only asymptotic estimates,
valid at sufficiemtly high $\eta$. To see their validity region we
present  in Fig. 6 the slopes $b$ in $\mu-\mu$ and
$p^2-\mu$ correlations as functions  of $\eta$ with $\mu_0=1$ calculated
by Monte-Carlo simulations, together with the
common asymptotic curve  (71). As one observes the asymptotic curve gets
more or less close to the exact ones  starting from $\eta>4\div 5$.
These are very high values from the physical point of view, attainable
only at energies in the multi-TeV region for the heaviest nuclei and
central collisions.

\section{References}

1. M.A.Braun and C.Pajares, Nucl. Phys. {\bf B390}(1993)542,549;\\
N.Armesto, M.A.Braun, E.G.Ferreiro and C.Pajares, Phys. Rev. Lett.
{\bf 77} (1996) 3736l\\
N.S.Amelin, M.A.Braun and C.Pajares, Phys.Lett. {\bf B 306} (1993) 312l
Z.Phys. {\bf C6 63} (1994) 507.

2. M.A.Braun, F.del Moral and C.Pajares, Phys. Rev. {\bf C65} (2002) 024907.

3. A.Capella and A.Krzywicki, Phys. Rev. {\bf D29} (1984) 1007;\\
P.Aurenche, F.Bopp and J.Ranft, Phys. Lett. {\bf B147} (1984) 212;\\
 A.Capella, J.Tran Thanh Van and J.Kwiecinski, Phys. Rev. Lett., {\bf 58}
(1987) 2015.

4. I.Derado {\it et al.}, Z.Phys. {\bf c 40} (1988) 25.

5. A.Capella and J. Tran Thanh Van, Phys. Rev., {\bf D 29} (1984) 2512.

6. T.T.Chou and C.N.Yang, Phys. Lett. {\bf B135} (1984) 175;
Phys. Rev., {\bf D 32} (1985) 1692.\\
M.A.Braun, C.Pajares and V.V.Vechernin, Phys. Lett., {\bf B 493} (2000)
54.

7. N.S.Amelin, N.Armesto, M.A.Braun, E.G.Ferreiro and C.Pajares,
Phys. Rev. Lett. {\bf 73} (1994) 2813

8. M.A.Braun and C.Pajares, Eur. Phys. J. {\bf C16} (2000) 349.

9. M.A.Braun, F.del Moral and C.Pajares, Eur. Phys. J. {\bf C21} (2001)
557.

10. V.V.Vechernin, R.S.Kolevatov, hep-ph/0304295; hep-ph/0305136.

11. M.A.Braun and C.Pajares,  Phys. Rev. Lett., {\bf 85} (2000) 4864.

\newpage
\section{Table}
{\bf Values of $\eta$ and average number of strings $<N>$
for central S-S and Pb-Pb collisions at different energies}

\begin{center}
{\bf S-S scattering ($b=0$)}\vspace{0.6cm}

\begin{tabular}{|r|r|r|}\hline
$\sqrt{s}$&$\eta$&$<N>$
\\\hline
19.4&0.40& 126\\\hline
62.5&0.52& 164\\\hline
 200&0.65& 204\\\hline
 546&0.80& 252\\\hline
1800&0.97& 308\\\hline
7000&1.23& 390\\\hline
\end{tabular}
\end{center}
\vspace*{0.6 cm}

\begin{center}
{\bf Pb-Pb scattering ($b=0$)}\vspace{0.8cm}

\begin{tabular}{|r|r|r|}\hline
$\sqrt{s}$&$\eta$&$N$
\\\hline
  19.4&1.08 &1190\\\hline
  62.5&1.39 &1536\\\hline
 200.0&1.60 &1800\\\hline
 546.0&2.03 &2240\\\hline
1800.0&2.46 &2720\\\hline
7000.0&3.03 &3350\\\hline
\end{tabular}
\end{center}

\newpage

\begin{figure}
\centerline{\epsfig{file=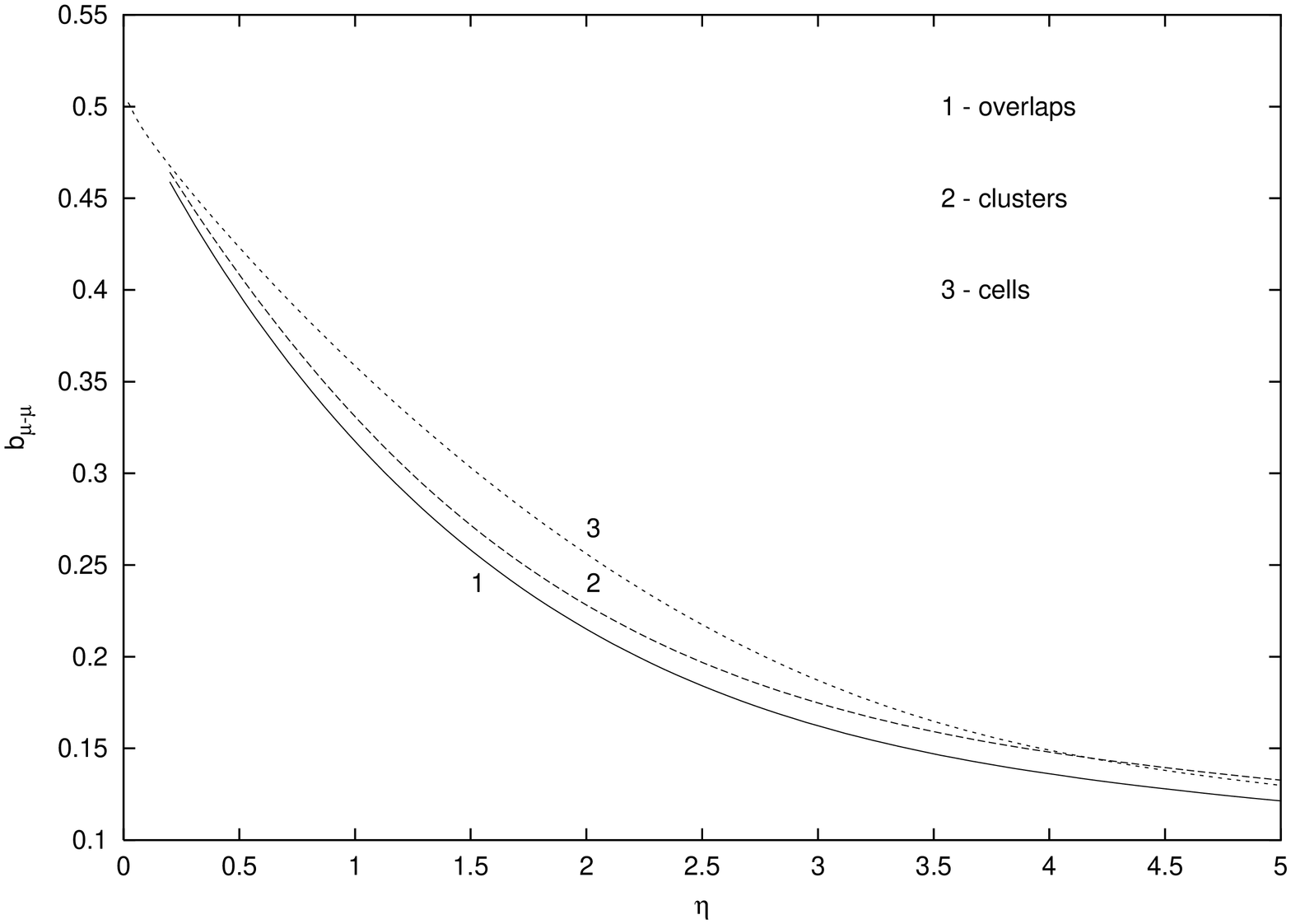,width=10cm,angle=-90}}
\caption[dummy]{
Correlation coefficient $b$ for $\mu-\mu$ correlations in the three
scenarios studied.
}
\end{figure}

\begin{figure}
\centerline{\epsfig{file=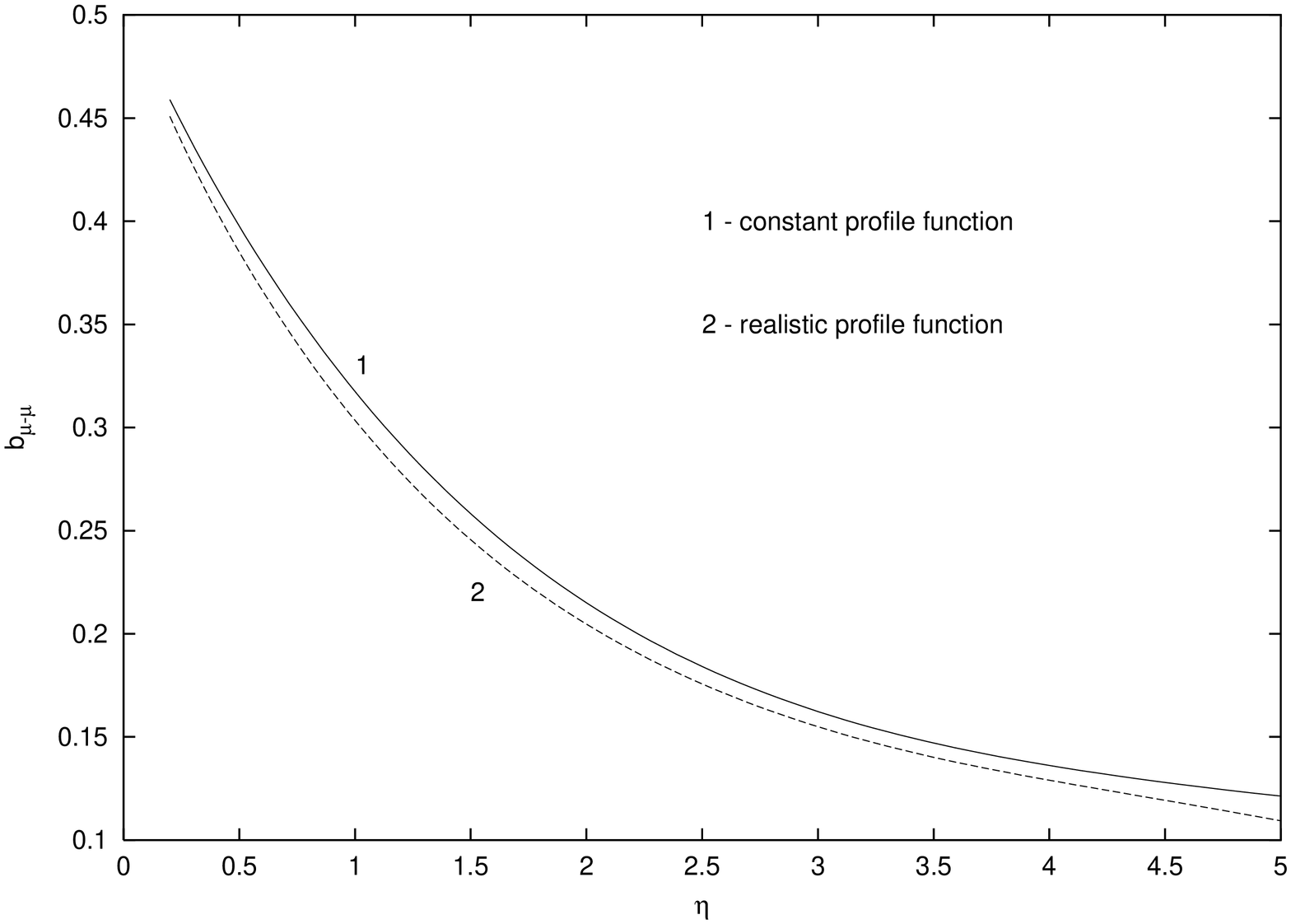,width=10cm,angle=-90}}
\caption[dummy]{
Correlation coefficient $b$ for $\mu-\mu$ correlations in the
overlap scenario for the homogeneous distribution of strings and the one
proportional to $T(b)\propto \sqrt{R_A^2-b^2}$.
}
\end{figure}

\begin{figure}
\centerline{\epsfig{file=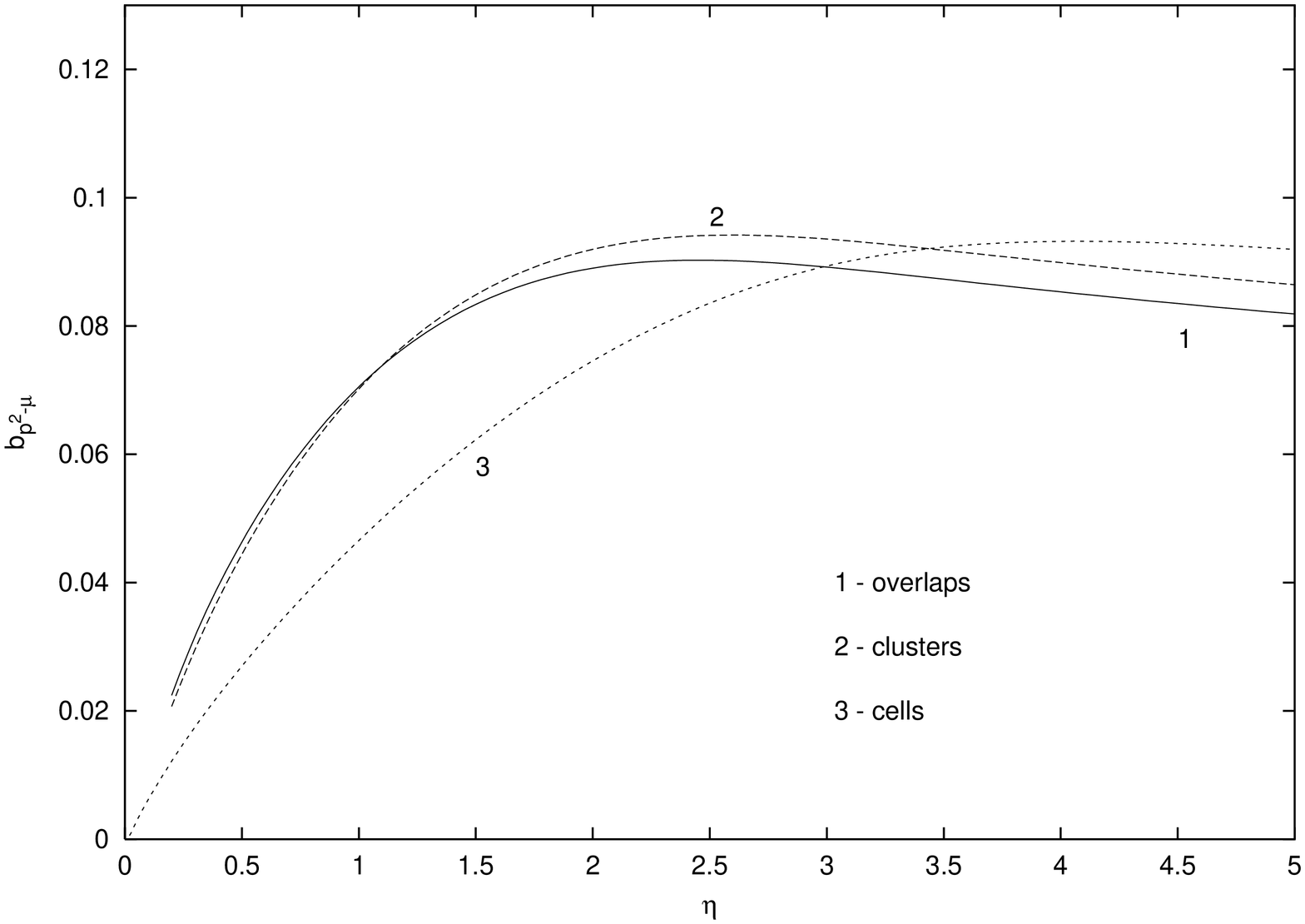,width=10cm,angle=-90}}
\caption[dummy]{
Correlation coefficient $b$ for $p^2-\mu$ correlations in the
three scenarios studied.
}
\end{figure}

\begin{figure}
\centerline{\epsfig{file=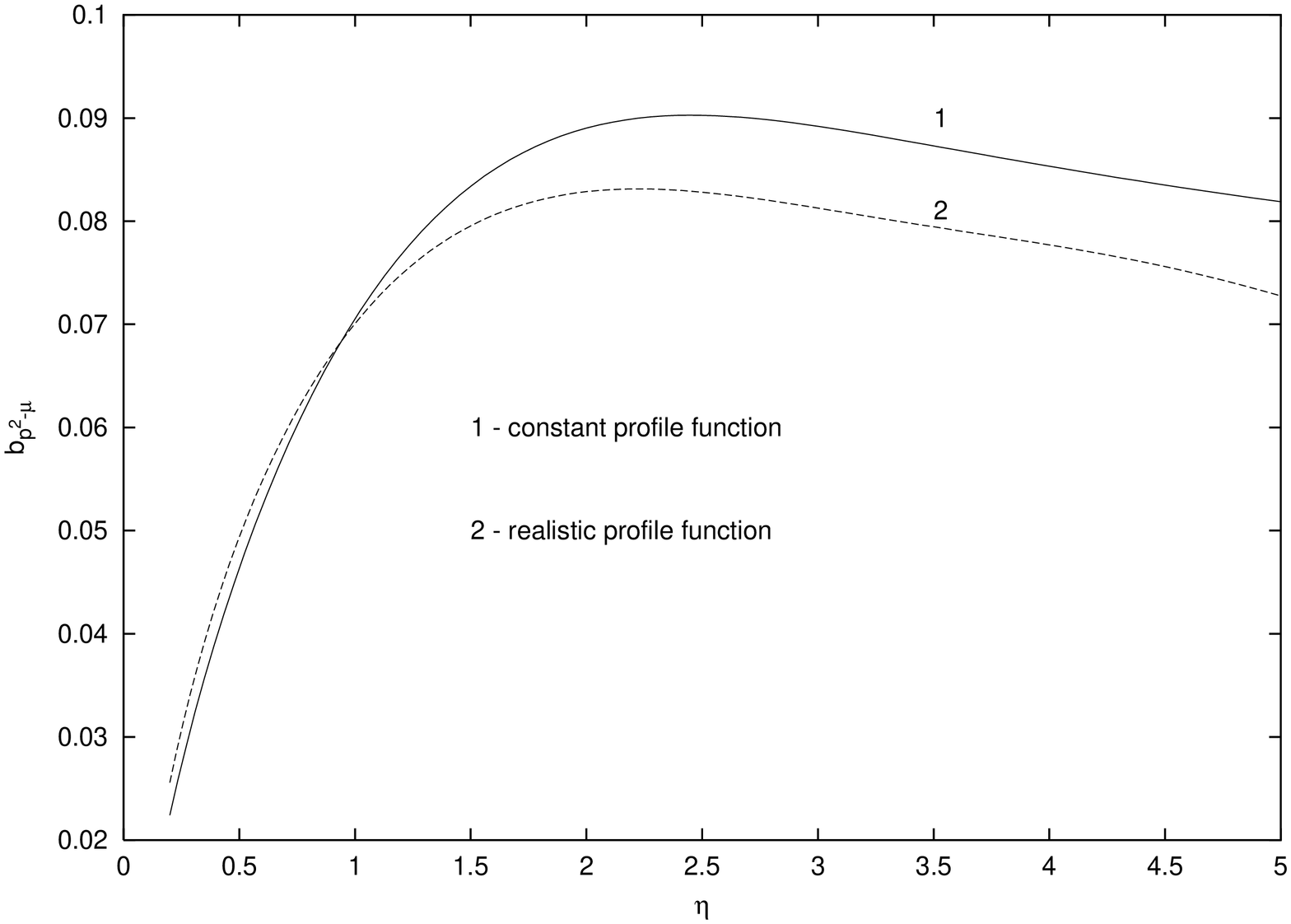,width=10cm,angle=-90}}
\caption[dummy]{
Correlation coefficient $b$ for $p^2-\mu$ correlations in the
overlap scenario for the homogeneous distribution of strings and the one
proportional to $T(b)\propto \sqrt{R_A^2-b^2}$.
}
\end{figure}

\begin{figure}
\centerline{\epsfig{file=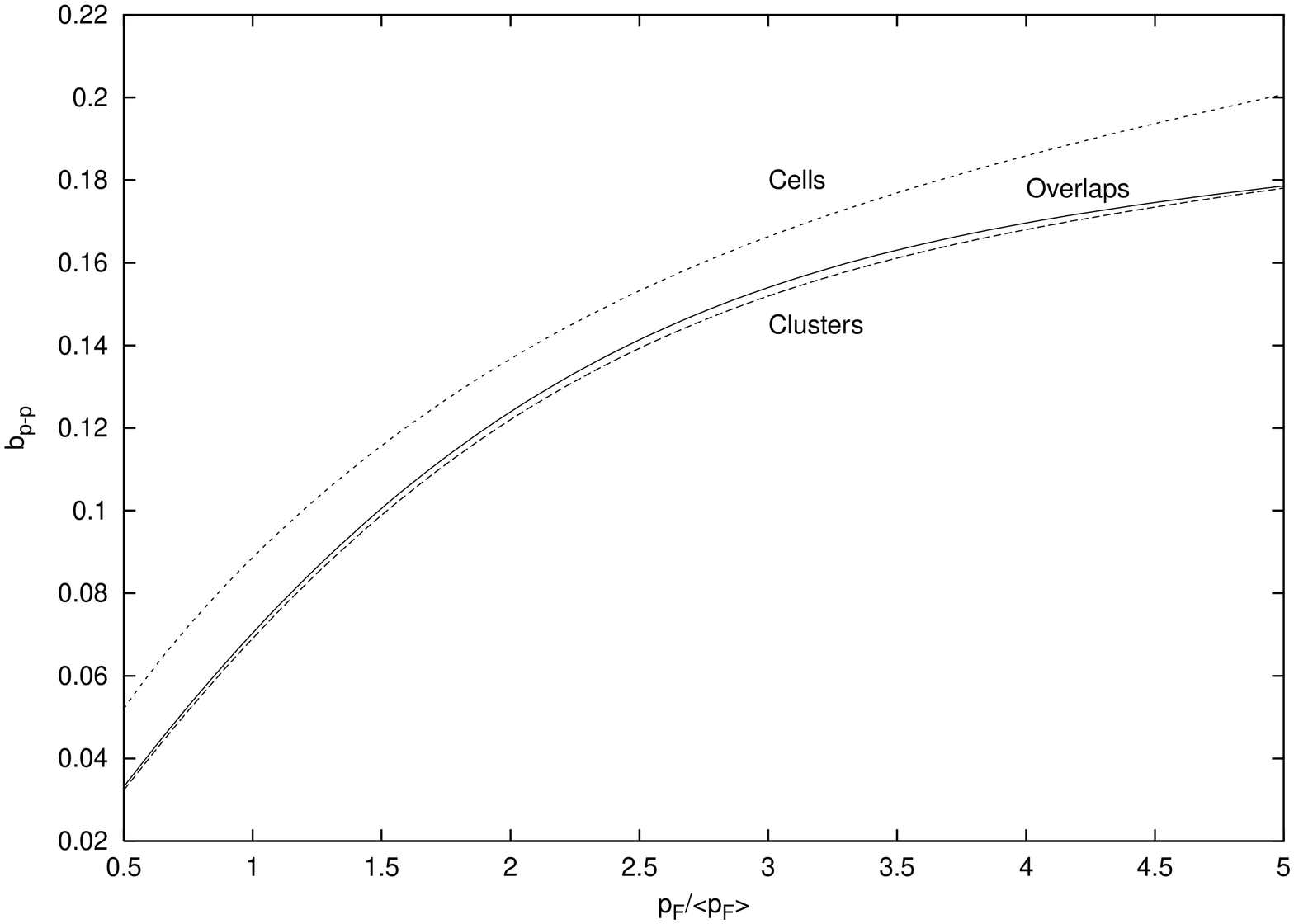,width=10cm,angle=-90}}
\caption[dummy]{
Correlation coefficient $b$ for $p-p$ correlations in the three
scenarios studied.
}
\end{figure}

\begin{figure}
\centerline{\epsfig{file=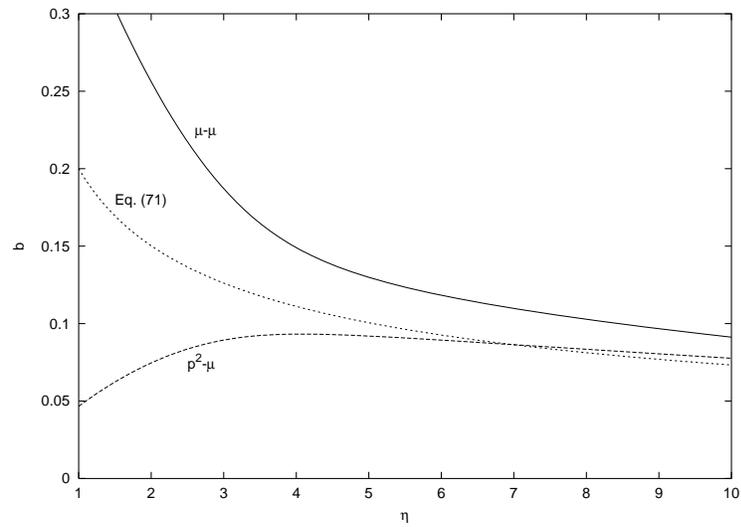,width=10cm,angle=-90}}
\caption[dummy]{
Correlation coefficients for $\mu-\mu$ and $p^2-\mu$ correlations
in the cell model together with the asymptotics (71).
}
\end{figure}

\end{document}